\documentclass[11pt,twocolumn,twoside]{article}
\usepackage{fully3d}

\usepackage{footnote} 
\usepackage{microtype}
\usepackage{graphicx}
\usepackage{amssymb}

\begin{document}

\title{Learning Wavelet-Sparse FDK for 3D Cone-Beam CT Reconstruction} 

\author[1]{Yipeng Sun}
\author[1]{Linda-Sophie Schneider}
\author[1]{Chengze Ye}
\author[1]{Mingxuan Gu}
\author[1]{Siyuan Mei}
\author[1]{Siming Bayer}
\author[1]{Andreas Maier}

\affil[1]{Pattern Recognition Lab, Friedrich-Alexander University Erlangen-Nuremberg, Erlangen, Germany}

\maketitle
\thispagestyle{fancy}

\begin{customabstract}
Cone-Beam Computed Tomography (CBCT) is essential in medical imaging, and the Feldkamp-Davis-Kress (FDK) algorithm is a popular choice for reconstruction due to its efficiency. However, FDK is susceptible to noise and artifacts. While recent deep learning methods offer improved image quality, they often increase computational complexity and lack the interpretability of traditional methods. In this paper, we introduce an enhanced FDK-based neural network that maintains the classical algorithm's interpretability by selectively integrating trainable elements into the cosine weighting and filtering stages. Recognizing the challenge of a large parameter space inherent in 3D CBCT data, we leverage wavelet transformations to create sparse representations of the cosine weights and filters. This strategic sparsification reduces the parameter count by $93.75\%$ without compromising performance, accelerates convergence, and importantly, maintains the inference computational cost equivalent to the classical FDK algorithm. Our method not only ensures volumetric consistency and boosts robustness to noise, but is also designed for straightforward integration into existing CT reconstruction pipelines. This presents a pragmatic enhancement that can benefit clinical applications, particularly in environments with computational limitations.
\end{customabstract}

\section{Introduction}
\label{sec1}

Cone-Beam Computed Tomography (CBCT) plays a crucial role in medical imaging by providing volumetric representations of the human body through the reconstruction of multiple X-ray projections \cite{schulze2011artefacts}. In the realm of CBCT reconstruction, Feldkamp-Davis-Kress (FDK) method are widely adopted due to their computational efficiency and ease of implementation \cite{feldkamp1984practical}. These characteristics make FDK the preferred choice in clinical settings, especially in hospitals equipped with lower-end CT machines. However, the FDK algorithm is sensitive to noise and artifacts, which can degrade the quality of the reconstructed images \cite{sun2024data, qi2024direct}.

In recent years, deep learning approaches have shown promise in improving CT reconstruction quality \cite{ye2024draco, ye2024deep, wurfl2016deep, maier2019learning}. While these methods can enhance image accuracy, they often introduce computational overhead and function as black boxes, compromising the interpretability of the reconstruction process. This lack of transparency is undesirable in clinical applications where understanding the underlying mechanisms is essential for diagnosis and treatment planning.

Inspired by our previous work in 2D parallel-beam CT reconstruction, which introduced trainable Fourier series for filter construction within the Filtered Backprojection (FBP) framework~\cite{sun2024data}, we extend the concept of known operator~\cite{maier2019learning} to 3D CBCT. In this work, we propose a modified FDK algorithm incorporating deep learning elements in a targeted manner. Specifically, we develop an FDK-based neural network where only the cosine weighting and filtering components are trainable, while the remaining aspects of the FDK algorithm are preserved. This selective integration of deep learning maintains much of the interpretability inherent to FDK while enabling data-driven optimization.

Given the 3D nature of CBCT data and the size of its detector array, the number of trainable parameters can become exceedingly large, often reaching several million. To address this issue, we employ wavelet transformations \cite{grossmann1984decomposition} to construct the cosine weights and filters, significantly reducing the parameter space. Through wavelet coefficient thresholding, we reduce the number of parameters by $93.75\%$ while maintaining performance. This sparsity not only accelerates convergence, but also retains the interpretability of classical algorithms while incorporating data-driven robustness.

Moreover, our approach adds no extra computational cost during inference compared to the classical FDK algorithm, which is important for practical clinical use. This efficiency makes it a helpful solution for healthcare facilities lacking advanced CT equipment. Additionally, our method is designed as a plug-and-play solution, facilitating seamless integration into more complex models that take sinograms as inputs and output reconstructed volumes. This adaptability allows for effortless enhancement and refinement of existing CT reconstruction frameworks.

This paper is organized as follows: Section~\ref{sec2} presents the theoretical foundation of our method, Section~\ref{sec3} describes the experimental setup, Section~\ref{sec4} showcases the results, and Section~\ref{sec5} concludes the paper and drops a discussion of future work.

\section{Methods}
\label{sec2}

The FDK algorithm is an extension of the 2D fan-beam FBP method, adapted for 3D cone-beam geometry by treating the 3D reconstruction as a series of tilted fan-beam reconstructions~\cite{feldkamp1984practical}. At its core, the FDK algorithm consists of two key steps: cosine weighting and filtering. Cosine weighting corrects for the geometric distortions introduced by the cone-beam configuration, while filtering removes excessive low-frequency components from the projections before backprojection.

To introduce trainable parameters into the FDK algorithm, we reformulate both the weighting and filtering steps using 2D wavelet transforms. Specifically, we employ a level-2 decomposition using the Daubechies wavelet basis~\cite{daubechies1988orthonormal}. By training only the approximation coefficients and setting all detail coefficients to zero, we enable the network to learn essential structural information.

\subsection{Choice of Wavelet Basis}
We use the Daubechies wavelet basis (db1), which is equivalent to the Haar wavelet, due to its efficiency, orthogonality, and symmetry properties \cite{daubechies1988orthonormal}. These characteristics make it particularly suitable for compactly representing 2D data while preserving essential features. Compared to other common wavelet bases such as Symlets or higher-order Daubechies wavelets, db1 is computationally less expensive and avoids unnecessary complexity in the representation. Additionally, its orthogonality ensures a lossless reconstruction.

The decomposition level (level-2) was chosen based on empirical results from preliminary experiments, which indicated that lower levels of decomposition did not yield significant improvements in performance but increased computational overhead. A level-2 decomposition strikes a balance between capturing essential features and maintaining efficiency.

\subsection{Trainable Cosine Weighting}

The cosine weighting factor \( w(s, v) \) is essential for compensating the divergence in cone-beam projections. It has dimensions matching the detector size, forming a 2D matrix \( w \in \mathbb{R}^{N_s \times N_v} \), where \( N_s \) and \( N_v \) represent the detector dimensions in the transverse (\( s \)) and axial (\( v \)) directions. Due to the circular orbits, this weighting matrix is identical for all projections.

We construct the trainable weighting factor using a 2D wavelet transform. First, we perform a level-2 decomposition of the initial weighting matrix \( w(s, v) \):

\begin{equation}
W = \mathcal{W}\{ w(s, v) \},
\end{equation}

where \( \mathcal{W} \) denotes the 2D wavelet transform,  $W$ is the wavelet coefficients matrix. We then focus on the approximation coefficients \( w_{LL} \) at level 2, which capture the most significant features of the weighting function. These approximation coefficients are set as trainable parameters. All detail coefficients, which represent finer-scale information, are set to zero to reduce computation load:

\begin{equation}
W = \begin{bmatrix}
 w_{LL}  & \mathbf{0} & \mathbf{0} & \mathbf{0} \\
\mathbf{0} & \mathbf{0} & \mathbf{0} & \mathbf{0} \\
\mathbf{0} & \mathbf{0} & \mathbf{0} & \mathbf{0} \\
\mathbf{0} & \mathbf{0} & \mathbf{0} & \mathbf{0}
\end{bmatrix}.
\label{eq:wavelet_weighting_coefficients}
\end{equation}

We then reconstruct the weighting factor \( w(s, v) \) using the inverse 2D wavelet transform $\mathcal{W}^{-1}$:

\begin{equation}
w(s, v) = \mathcal{W}^{-1}\{ W \}.
\label{eq:wavelet_weighting_reconstruction}
\end{equation}

This approach allows the network to learn an optimal weighting function by adjusting the approximation coefficients \( w_{LL} \) during training efficiently.

The weighted projection \( p_w(\theta, s, v) \) is then computed by element-wise multiplication of the reconstructed weighting factor and the original projection data \( p(\theta, s, v) \):

\begin{equation}
p_w(\theta, s, v) = w(s, v) \cdot p(\theta, s, v),
\label{eq:weighted_projection}
\end{equation}

where \( \theta \) represents the projection angle.

\subsection{Trainable Filtering}

Traditionally, each projection in cone-beam CT uses a 1D filter along the transverse  (\( s \))-axis, applied uniformly across the axial (\( v \))-axis. In our method, we introduce trainable parameters into the filtering step by utilizing a 2D wavelet transform on the set of filters for all projections.

We consider the collection of 1D filters for all projections as a 2D matrix \( h \in \mathbb{R}^{M \times N_s} \), where \( M \) is the number of projections and \( N_s \) is the detector width. Each row \( h_m(s) \) corresponds to the filter for projection \( \theta_m \).

We perform a 2D wavelet decomposition on this filter matrix:

\begin{equation}
H = \mathcal{W}\{ h(m, s) \},
\end{equation}

where \( \mathcal{W} \) denotes the 2D wavelet transform along the projection index \( m \) and the detector width \( s \). We set the approximation coefficients \( f_{LL} \) at the coarsest level as trainable parameters and set all detail coefficients to zero:

\begin{equation}
H = \begin{bmatrix}
 f_{LL} & \mathbf{0} & \mathbf{0} & \mathbf{0} \\
\mathbf{0}         & \mathbf{0} & \mathbf{0} & \mathbf{0} \\
\mathbf{0}         & \mathbf{0} & \mathbf{0} & \mathbf{0} \\
\mathbf{0}         & \mathbf{0} & \mathbf{0} & \mathbf{0}
\end{bmatrix}.
\label{eq:wavelet_filter_coefficients}
\end{equation}

We reconstruct the filter matrix using the inverse 2D wavelet transform:

\begin{equation}
h(m, s) = \mathcal{W}^{-1}\{ H \}.
\label{eq:wavelet_filter_reconstruction}
\end{equation}

This yields a set of 1D filters \( h_m(s) \) for each projection at \( \theta_m \). Each filter \( h_m(s) \) is then applied to the corresponding weighted projection \( p_w(\theta_m, s, v) \) by convolving along the \( s \)-axis:

\begin{equation}
p_f(\theta_m, s, v) = p_w(\theta_m, s, v) \ast_s h_m(s),
\label{eq:filtered_projection}
\end{equation}

where \( \ast_s \) denotes convolution along the \( s \)-axis.

Alternatively, in the Fourier domain, the convolution becomes an element-wise multiplication:

\begin{equation}
p_f(\theta_m, s, v) = \mathcal{F}_s^{-1}\left\{ \mathcal{F}_s\{ p_w(\theta_m, s, v) \} \cdot H_m(s) \right\},
\label{eq:filtered_projection_fourier}
\end{equation}

where \( \mathcal{F}_s \) and \( \mathcal{F}_s^{-1} \) denote the Fourier and inverse Fourier transforms along the \( s \)-axis, and \( H_m(s) = \mathcal{F}_s\{ h_m(s) \} \).

By treating the collection of filters as a 2D matrix and performing a 2D wavelet transform, we allow the network to learn correlations between filters across different projections. Training only the approximation coefficients at the coarsest level enables the filters to capture essential features.

\subsection{Integration into a Neural Network}

The final reconstructed image \( f(x, y, z) \) is computed by backprojecting the filtered projections over all projection angles:

\begin{equation}
f(x, y, z) = \sum_{m=0}^{M-1} p_f\left( \theta_m, s_{xyz}^m, v_{xyz}^m \right) \Delta \theta,
\label{eq:reconstruction}
\end{equation}

where \( s_{xyz}^m \) and \( v_{xyz}^m \) are the detector coordinates corresponding to the voxel at \( (x, y, z) \) for projection angle \( \theta_m \), and \( \Delta \theta \) is the angular increment between projections.

\begin{figure}
\centering
\includegraphics[width=1.0\linewidth]{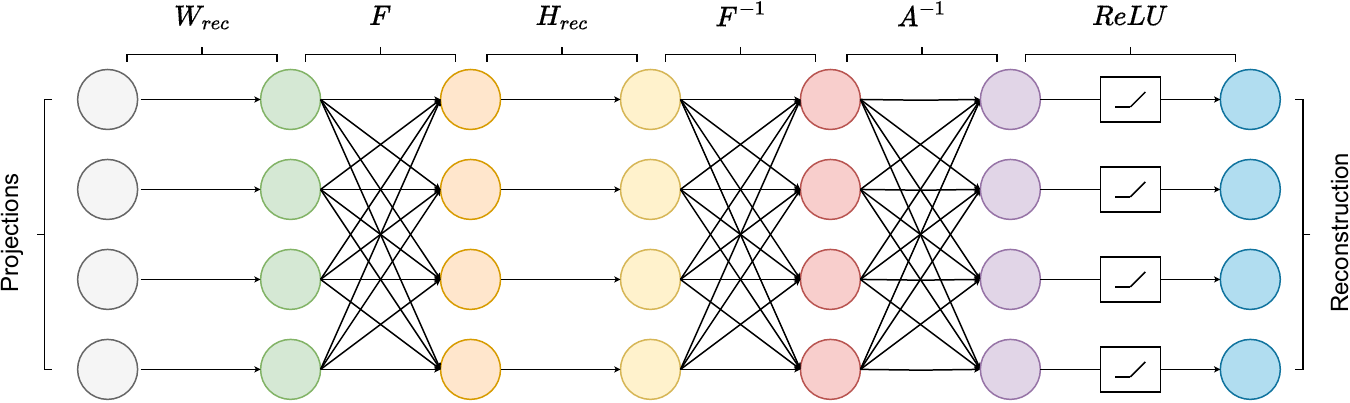}
\caption{Network for $I_{\text{rec}} = \text{ReLU}( A^{-1} {F}^{-1}H_{rec}{F} W_{rec}P )$ from projections $P$ to reconstruction ${I}_{rec}$.}
\label{fig:network}
\end{figure}

Following the principles of deep learning in computed tomography~\cite{maier2019learning}, we integrate the complete reconstruction process into a neural network architecture (illustrated in Fig \ref{fig:network}). To express the filtering operation in the Fourier domain while maintaining notational simplicity, we reformulate the reconstruction equation with condensed expressions.

Let \( W_{rec} = \mathcal{W}^{-1}\left\{ W_{\text{train}} \right\} \) represent the reconstructed weighting factor derived from the trainable wavelet approximate coefficient \(W_\text{train}\), and \( H_{rec} = \mathcal{W}^{-1}\left\{ H_{\text{train}} \right\} \) represent the reconstructed filter kernels derived from the trainable wavelet approximate coefficient \(H_\text{train}\) in the Fourier domain. Then, the reconstructed image \( I_{\text{rec}} \) can be expressed as:

\begin{equation}
I_{\text{rec}} = \text{ReLU}( A^{-1} {F}^{-1}H_{rec}{F} W_{rec}P ),
\label{eq:trainable_fdk_nn_fourier_compact}
\end{equation}

where:

\begin{itemize}
\item \( P \in \mathbb{R}^{M \times N_s \times N_v} \) is the original projection data.
\item \( W_{rec} \) is the reconstructed weighting factor obtained from the inverse 2D wavelet transform of the trainable coefficients \( W_{\text{train}} \).
\item \( H_{rec} \) is the reconstructed filter kernels, where the filter kernels are obtained from the inverse 2D wavelet transform of the trainable coefficients \( H_{\text{train}} \).
\item \( {F} \) and \( {F}^{-1} \) denote the Fourier and inverse Fourier transforms along the \( s \)-axis.
\item \( A^{-1} \) represents the cone-beam backprojection operator.
\item \( \text{ReLU} \) enforces non-negativity in the reconstruction.
\end{itemize}

\section{Experiments}
\label{sec3}

We conducted experiments using a simulated noisy dataset derived from the pancreatic cancer dataset provided by Hong et al. \cite{hong2021breath}. Forward projections were generated using the PYRO-NN framework \cite{syben2019pyronn}, with Poisson noise added to simulate realistic imaging conditions. The noise generation process followed the methodology described in \cite{mccollough2020low}. The dataset consists of abdominal CT scans from 40 patients, divided into $28$ scans for training, $6$ for validation, and $6$ for testing. The key CBCT geometry parameters for the dataset generation process are summarized in Table \ref{tab:parameters}.

\begin{table}[h]
\centering
\resizebox{\linewidth}{!}{%
\begin{tabular}{@{}cccc@{}}
\toprule
\textbf{Volume shape} & \textbf{Volume spacing} & \textbf{Number of projections} \\ 
\midrule
$512\times512\times512$ & $0.5 \, \mathrm{mm} \times 0.5 \, \mathrm{mm} \times 0.5 \, \mathrm{mm}$ & $400$ \\
\bottomrule
\toprule
\textbf{Angular range} & \textbf{Source-isocenter distance} & \textbf{Source-detector distance} \\
\midrule
$2\pi$ & $1200 \, \mathrm{mm}$ & $1500 \, \mathrm{mm}$ \\
\bottomrule
\toprule
\textbf{Detector shape} & \textbf{Detector spacing} \\
\midrule
$800 \times 800$ & $0.5 \, \mathrm{mm} \times 0.5 \, \mathrm{mm}$ \\
\bottomrule
\end{tabular}%
}
\caption{CBCT geometry parameters used for dataset generation, based on a circular orbit configuration.}
\label{tab:parameters}
\end{table}

All experiments were performed using a single Nvidia A6000 GPU with Python 3.11 and PyTorch 2.1.1. The training procedure involved 100 epochs, utilizing the Adam optimizer with a learning rate of $0.001$.

\section{Results}
\label{sec4}

\begin{figure}[ht]
\centering
\includegraphics[width=1.0\linewidth]{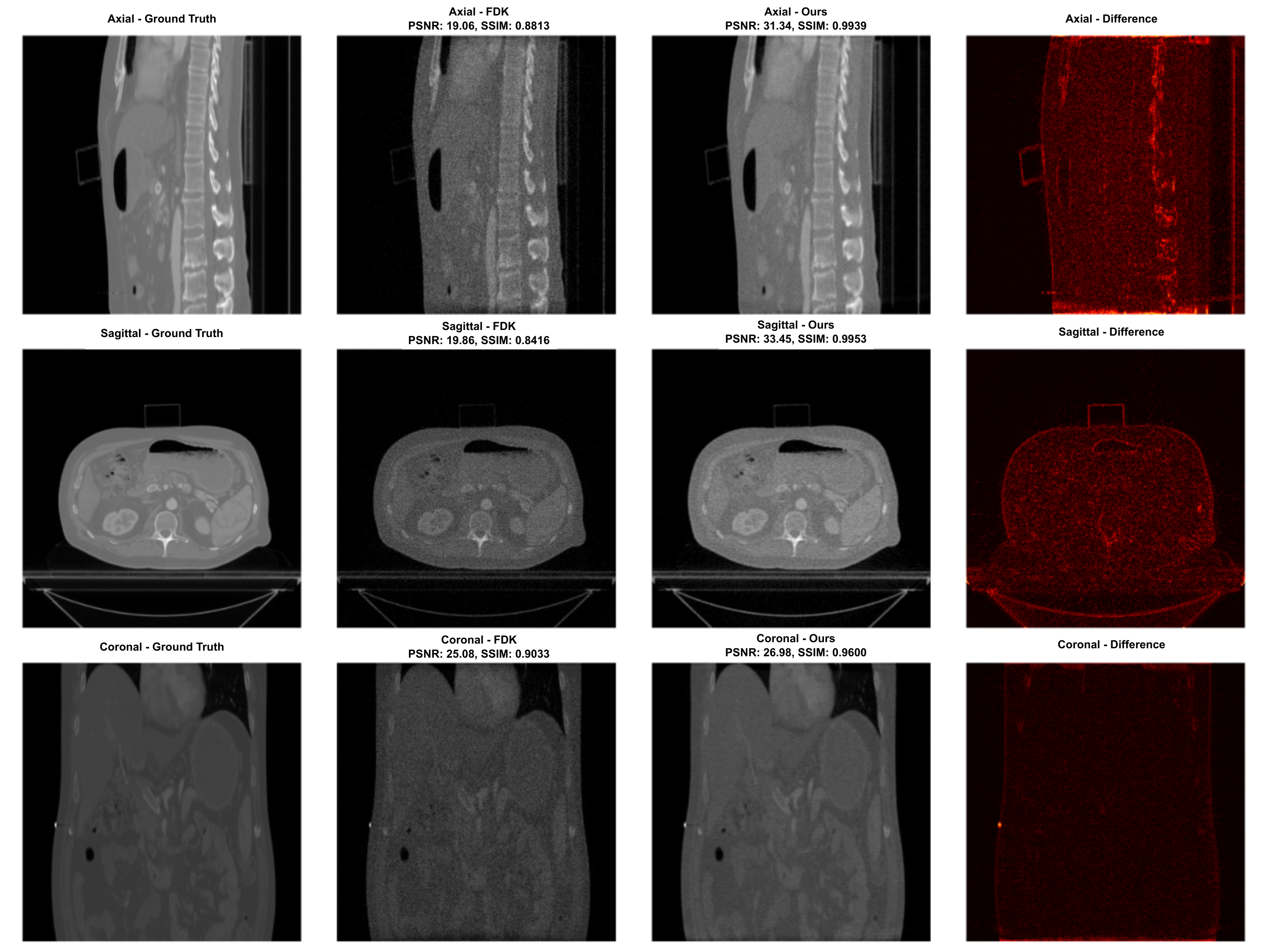}
\caption{Comparison of visualization results between our proposed method and FDK reconstruction (Ram-Lak filter). The figure shows axial, sagittal, and coronal views with their corresponding difference maps (Patient ID: 07).}
\label{fig:resluts}
\end{figure}

The results of our experiments demonstrate the effectiveness of the proposed method in improving reconstruction quality compared to the FDK algorithm. As shown in Figure \ref{fig:resluts}, the visualizations of Axial, Sagittal, and Coronal views highlight significant improvements in noise reduction and structural detail preservation achieved by our method. The difference images further emphasize the reduced reconstruction error relative to the ground truth. 

Quantitatively, as shown in Table \ref{tab:results}, our method consistently outperforms FDK across all views, demonstrating higher Peak Signal-to-Noise Ratio (PSNR) and Structural Similarity Index Measure (SSIM). In the axial view, for instance, our method achieves a PSNR of 31.94 dB and an SSIM of 0.9925, compared to FDK's 22.69 dB and 0.9368, respectively. These results demonstrate the superior robustness of our method to noise and enhanced capability for accurate reconstruction.

\begin{table}[ht]
\centering
\resizebox{0.65\linewidth}{!}{%
\begin{tabular}{lcc}
\toprule
\textbf{View}              & \textbf{PSNR (dB)$\uparrow$}         & \textbf{SSIM$\uparrow$}          \\ 
\midrule
Axial View (Ours)          & \textbf{31.94}        & \textbf{0.9925}        \\
Axial View (FDK)           & 22.69                & 0.9368                \\
Sagittal View (Ours)       & \textbf{30.12}        & \textbf{0.9614}        \\
Sagittal View (FDK)        & 24.91                & 0.9290                \\
Coronal View (Ours)        & \textbf{28.04}        & \textbf{0.9839}        \\
Coronal View (FDK)         & 21.84                & 0.9301                \\ 
\bottomrule
\end{tabular}%
}
\caption{Comparison of PSNR and SSIM metrics between our proposed method and the FDK algorithm across different views.}
\label{tab:results}
\end{table}

\section{Discussion and Conclusion}
\label{sec5}

In this work, we presented a novel FDK-based neural network for efficient 3D Cone-Beam CT reconstruction that integrates trainable wavelet-sparse weights and filters. By employing wavelet transformations, we significantly reduced the parameter space by 93.75\%, enabling faster convergence and maintaining interpretability. Our method retains the computational efficiency of the classical FDK algorithm during inference while enhancing robustness to noise and preserving volumetric consistency. Experimental results demonstrated that the proposed approach outperforms the standard FDK algorithm in terms of both quantitative metrics (PSNR and SSIM) and qualitative visual quality, making it a practical solution for clinical applications.

This work demonstrates the feasibility of incorporating trainable components into traditional algorithms, achieving enhanced performance while maintaining efficiency and interpretability. Future research will explore extending this framework to non-circular orbits and advanced imaging geometries, as well as integrating it with iterative reconstruction techniques to further improve image quality in challenging scenarios.

Despite its promising results, our method has limitations that warrant further investigation. First, while the wavelet sparsity assumption enables efficient parameter reduction, it may limit the ability of the network to adapt to highly irregular noise or artifact patterns. Second, our approach was tested on simulated datasets; its generalization to real-world clinical data, with more complex noise characteristics and diverse patient anatomies, remains to be validated. Additionally, the reliance on pre-defined wavelet bases could be optimized by learning these bases during training.

From a clinical perspective, computational efficiency during inference is crucial for widespread adoption, especially in resource-constrained settings. However, future work should also address the feasibility of real-time processing for larger datasets and more complex geometries, such as non-circular orbits. Moreover, integrating this framework with iterative reconstruction techniques could further improve image quality in cases with sparse data or incomplete projections.

\printbibliography
\end{document}